\documentclass{llncs}

\usepackage{hyperref}
\usepackage{graphicx}
\usepackage{rotating}
\usepackage{paralist}
\usepackage{footmisc}
\usepackage{booktabs}
 \usepackage{subfig}
 \usepackage[inline]{enumitem}
\usepackage{todonotes}
\usepackage[misc,geometry]{ifsym}

\usepackage{tabularx}
\newcolumntype{R}{>{\raggedleft\arraybackslash}X}

\usepackage{xcolor}

\usepackage{tikz}
\usetikzlibrary{arrows,shapes.geometric,positioning,matrix}
\usepackage{tikz-qtree}
\tikzset{
>=latex,
  ishikawa/.style={align=center, inner sep=0pt},
  matter/.style  ={rectangle, minimum size=6mm, very thick, draw=black,
     font=\itshape},
  level_1/.style ={rectangle, node distance=60pt, minimum size=6mm, very thick,
    font=\itshape},
  level_2/.style={rectangle, minimum size=6mm, font=\itshape, font=\tiny}}
\tikzset{
  rows/.style 2 args={@/.style={row ##1/.style={#2}},@/.list={#1}},
  cols/.style 2 args={@/.style={column ##1/.style={#2}},@/.list={#1}},
}

\begin{document}

\title{On the Distinction of Functional and Quality Requirements in Practice}

\author{Jonas Eckhardt\inst{1}\textsuperscript{\Letter} \and Andreas Vogelsang\inst{2} \and Daniel M\'endez Fern\'andez\inst{1}}

\institute{Technical University of Munich \\
\email{\{eckharjo,mendezfe\}@in.tum.de}
\and
Technische Universit\"at Berlin\\
\email{andreas.vogelsang@tu-berlin.de}}

\maketitle

\begin{abstract}
Requirements are often divided into functional requirements (FRs) and quality requirements (QRs). 
However, we still have little knowledge about to which extent this distinction makes sense from a practical perspective. 
In this paper, we report on a survey we conducted with 103 practitioners to explore whether and, if so, why they handle requirements labeled as FRs differently from those labeled as QRs. We additionally asked for consequences of this distinction w.r.t. the development process. 
Our results indicate that the development process for requirements of the two classes strongly differs (e.g., in testing). We identified a number of reasons why practitioners do (or do not) distinguish between QRs and FRs in their documentation and we analyzed both problems and benefits that arise from that. We found, for instance, that many reasons are based on expectations rather than on evidence. Those expectations are, in fact, not reflected in specific negative or positive consequences per se. It therefore seems more important that the decision whether to make an explicit distinction or not should be made consciously such that people are also aware of the risks that this distinction bears so that they may take appropriate countermeasures.
\vspace{-1em}
\keywords{Quality requirements, functional requirements, survey}
\vspace{-1em}
\end{abstract}

\section{Introduction}
In literature (e.g.,~\cite{IEEE29148,Pohl10,robertson2012mastering,sommerville1998requirements,van2001goal}), requirements are often categorized in {\itshape functional requirements (FRs)}, {\itshape quality requirements (QRs)}, and {\itshape constraints}. FRs are characterized as ``things the product must do'' contrasting QRs as ``qualities the product must have''  and constraints as ``organizational or technological requirement''. Although this categorization is common sense to some degree, there are still debates about the precision of the categories (e.g.,~\cite{Glinz07}). There are other academic groups that suggest to rather distinguish between {\itshape behavior} (e.g., response times) and {\itshape representation} (e.g., programming languages)~\cite{Broy16Rethinking}.

In a previously conducted study~\cite{Eckhardt16}, we analyzed 11 requirements specifications from industrial environments with a particular focus on requirements labeled as ``quality''. We found out that (i) there is a distinction between QRs and FRs in the documentations, and that (ii) many requirements labeled as QR actually describe system behavior and, thus, could also be labeled as FR. However, our previous investigation focused on analyzing artifacts after the fact and we still have little knowledge about what difference it makes in a development process if a requirement is labeled as FR or as QR and what the resulting consequences are. In response to this question, we conducted a survey with 103 practitioners which we report in this paper. 

In particular, we contribute:
\begin{enumerate*}[label={(\roman*)}]
\item a quantification of company practices regarding the style of documenting functional and quality requirements, 
\item a list of reasons why practitioner do or do not document FRs and QRs separately,
\item a list of consequences for the two styles of documentation that helps engineers to make conscious decisions. 
\end{enumerate*}


%

\section{Research Objective}
\label{sect:objective}
The goal of this study is to understand whether practitioners consider product-related requirements labeled as FR differently from those labeled as QR.
We are further interested in the reasons for this distinction and the resulting consequences for the development process. 
We derive the following research questions:\looseness=-1

\noindent {\bfseries RQ1: Do practitioners handle FRs and QRs differently?} In this RQ, we want to analyze whether QRs are documented in practice, whether there is a distinction in the documentation, and whether this distinction makes a difference in the development process. To this end, we formulate the following sub-RQs:
\begin{compactitem}
\item[{\bfseries RQ1.1}] {\bfseries Do practitioners differentiate between QRs and FRs in the documentation?} 
We want to know whether the accepted categorization of product-related requirements as FRs or QRs is reflected in the style of documentation as used in practice.
\item[{\bfseries RQ1.2}] {\bfseries To what extent do development activities for QRs differ from activities for FRs?}
A possible consequence of a requirement categorization is that different categories of requirements are handle differently in the development process. We want to investigate whether this is the case in practice and how this is influenced by the style of documentation.
\end{compactitem} 
\noindent {\bfseries RQ2: What are reasons for distinguishing or not distinguishing between QRs and FRs in the documentation?} 
While categorizations only provide definitions, we are interested in the underlying reasons that lead practitioners to distinguish or not distinguish between QRs and FRs in the documentation. \looseness=-1

\noindent {\bfseries RQ3: What are positive and negative consequences of distinguishing or not distinguishing QRs and FRs in the documentation?}
A decision for or against a separate documentation may have positive or negative consequences that practitioners should be aware of.

\section{Research Methodology}
\label{sect:method}
Our goal was to reach out to a broad spectrum of practitioners and capture their perceptions of their own project environments. To this end, we used (online) survey research as our main vehicle. We intentionally designed the survey such that respondents required as little effort as possible to complete it; we kept the number of questions at a minimum, the instrument was self-contained and it included all relevant information. We further limited the response types to numerical, Likert-scale, and short free form answers as suggested by Kitchenham and Pfleeger~\cite{Kitchenham08}. As a validation of our instrument and its alignment with the audience, we piloted the survey with three practitioners, who completed the survey and afterwards participated in an interview, where questions and answers where checked for misunderstandings.


\subsection{Subject Selection}
We deliberately targeted practitioners who work with requirements. This includes practitioners who write requirements (e.g., \emph{requirements engineers}) but also practitioners whose work is based on requirements (e.g., \emph{developers} or \emph{testers}), and also practitioners who manage projects or requirements.
Our survey was further conducted anonymously. Since we were not able to exactly control who is answering the survey, it was especially important to follow Kitchenham and Pfleeger's~\cite{Kitchenham08} advice on the need to understand whether the respondents had enough knowledge to answer the questions in an appropriate manner. For this, we excluded data from respondents who answered that they do not use requirements specifications at all, or respondents who stated that they did not know how requirements are handled in their company. 
We finally offered respondents the chance to leave an email address if they were interested in the results of the survey.

\subsection{Data Collection and Instrument}
We started our data collection on February 4th, 2016 and closed the survey on February 22nd, 2016. For inviting practitioners to participate, we did not select a specific closed group of practitioners but, instead, contacted as many practitioners as possible via the authors' personal contacts from previous collaborations, via public mailing lists such as \emph{RE-online}, and via social networks. 
In the following, we introduce the main elements of our instrument used. The full instrument can be taken from our online material\footnote{\url{http://www4.in.tum.de/~eckharjo/SurveyResults.zip}\label{ftn:onlineMaterial}}. 

\noindent {\bfseries Demographics: }
We collected a set of demographic data from the respondents to interpret and triangulate the data with respect to different contexts of the respondents. The demographic data included the role of the participant, the experience, the company's size, the typical project size, the geographical distribution of project members, the paradigm of their applied development process (on a scale from agile to plan-driven), the industrial sector, the type of developed systems, and the role of the requirements specification within the company.
To better understand the participant's focus and project context, we additionally asked  respondents for the importance of different types of QRs in their projects. The respondents were asked to assess the importance of quality factors\footnote{These were functional suitability, performance\slash efficiency, compatibility, usability, reliability, security, maintainability, and portability.} taken from ISO\slash IEC~25010~\cite{iso25010} for their typical projects on a 5-point Likert scale. 

\noindent {\bfseries Practices of Handling QRs: }
As a first step towards comparing different practices for handling QRs, we asked the respondents how strongly development activities differ between QRs and FRs in the phases \emph{requirements engineering}, \emph{architecture\slash design}, \emph{implementation}, and \emph{testing}. As a follow up, we provided a free form text field and asked the respondents to explain the differences in detail.

\begin{figure}
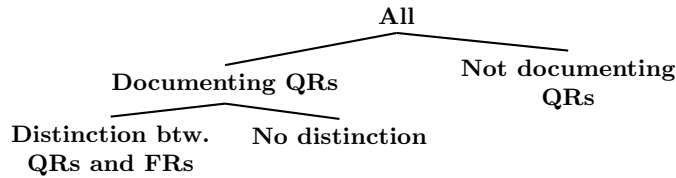

\centering
\footnotesize
\tikzset{edge from parent/.append style={thick}}
\tikzset{level distance=25pt,sibling distance=-5pt}
\Tree [.{\bfseries All}  [.{\bfseries \parbox{4cm}{\centering Documenting QRs}} [.{\bfseries \parbox{3cm}{\centering Distinction btw. QRs and FRs}} ] [.{\bfseries \parbox{3cm}{\centering No distinction\\{\color{white}{a}}}} ] ] [.{\bfseries \parbox{3cm}{\centering Not documenting QRs}} ] ]
  \caption{Categorization of respondents by their style of documenting QRs.}
  \label{fig:participantgroups}
\end{figure}

We were especially interested in the question whether it makes a difference for the development process if project participants distinguish between QRs and FRs and how this distinction is documented. Therefore, we asked the respondents two conditional questions. First, we asked whether QRs are explicitly documented in their projects. If this was the case, we asked whether the respondents explicitly distinguish between QRs and FRs in the documentation, i.e. whether they are labeled differently (e.g., some requirements are labeled as \emph{performance} or \emph{maintainability}) or documented in different sections (e.g., special sections for \emph{performance} or \emph{maintainability}). The answers to these questions categorize the responses into three groups (see also \figurename~\ref{fig:participantgroups}).

\noindent {\bfseries Problems\slash Benefits of Current Practices: }
Given the categorization into the three groups, we asked our respondents for specific reasons why they do or do not distinguish between QRs and FRs. Additionally, we asked for benefits and problems that arise from the way they consider QRs (i.e., not documenting QRs, mixing QRs and FRs in the documentation, or distinguishing between QRs and FRs in the documentation). For these questions, we provided free form text fields to be filled out by the respondents.

\subsection{Data Analysis}
Our data analysis constitutes a mix of descriptive statistics and qualitative text analysis. To answer RQ1, we analyzed in particular the answers that the respondents provided for the following survey questions:
\begin{enumerate*}[label={(\roman*)}]
\item \emph{Are QRs documented in your typical projects},
\item \emph{In the documentation (e.g., in a requirements specification), do you distinguish between QRs and FRs},
\item \emph{Considering the following phases, how much do the activities for handling QRs differ from those for FRs}, and
\item \emph{Considering your work, for what activities does it make a difference if you consider an QRs vs. an FR}.
\end{enumerate*}
For RQ1.1 and RQ1.2 we analyzed the results of the first, second, and third question, respectively. As the answers for the fourth question are open, we analyzed the answers in detail to provide more insights in the activities and the differences. 

To answer RQ2 and RQ3, we analyzed the data our respondents provided for the following survey questions:
\begin{enumerate*}[label={(\roman*)}]
\item \emph{Are there specific reasons why you do (or do not) distinguish between QRs and FRs in the documentation},
\item \emph{Do you experience negative consequences in your current work that result from distinguishing (not distinguishing) between QRs and FRs in the documentation}, and
\item \emph{Do you experience positive consequences in your current work that result from distinguishing (not distinguishing) between QRs and FRs in the documentation}.
\end{enumerate*}
The answers to the questions are free text answers. To analyze the results, we coded the provided answers in pairs of researchers to assemble a conceptual model of reasons and consequences for distinguishing between QRs and FRs in practice. The qualitative coding technique was chosen as recommended by (Straussian) Grounded Theory~\cite{SRF16}, but differs in that the central categories were previously defined following our research questions. To visualize our results from the text analysis, we used cause-effect diagrams (also known as Ishikawa diagrams).

\section{Study Results}
\label{sect:results}

\subsection{Sample Characterization}


In total, 283 people clicked on the link to our survey, 172 started the survey (61\%), and 109 completed it (39\%). From these 109 respondents, we excluded 6 as they matched our exclusion criteria. The respondents seem quite experienced as 93\% stated that they have more than 3 years of experience with requirements, 5\% one to three years, and only 2\% with less than a year. Furthermore, a majority of the respondents work in large companies: 57\% work in companies with more than 2000 employees, 25\% in companies with 250--2000 employees, and 17\% in companies with less than 250 employees. However, typical projects of the respondents showed a variety of small to large projects: 24\% stated that in a usual project in their company up to 10 people are involved, 46\% that 11--50 people are involved, 24\% that more than 50 people are involved, and 6\% did not know. Most of the respondents (59\%) answered that their team is distributed over multiple locations in more than one country, 23\% that the team is distributed over multiple locations but in one country, and 17\% that all team members are in one location. The employed process paradigm is balanced between agile and plan-driven: 41\% of the respondents answered that their development process is rather agile, 21\% that it is rather plan-driven, 37\% that it is mixed, and 1\% did not know. 
The type of systems the respondents develop is quite balanced (except for consumer software): 24\% develop embedded systems, 37\% business information systems,  5\% consumer software, and 34\% hybrid systems.
Most of the respondents use requirements specifications for in-house development (57\%), 23\% create them and an external company is responsible for the development, and 19\% are subcontractors using requirements specifications (e.g., as basis for development or testing). 
\looseness=-1

\subsection{RQ1: Handling of QRs in practice} 
\begin{figure}
\centering
      \subfloat[Do you document QRs?]{
      \includegraphics[width=0.45\columnwidth]{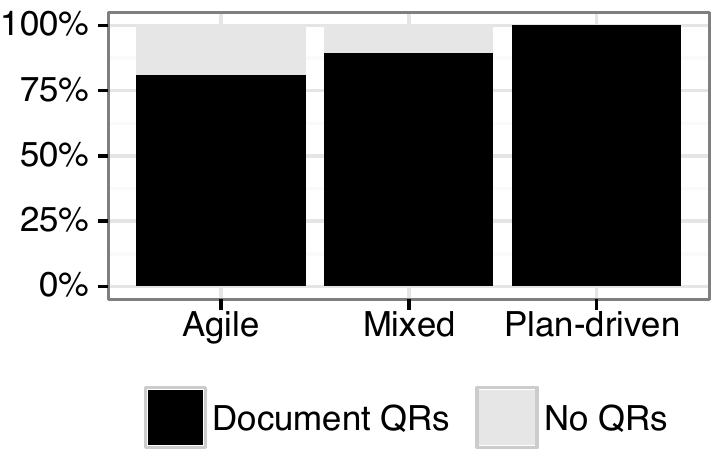}
	\label{fig:agilityvsdocumented1}
	}
	\hspace{0.2cm}
      \subfloat[Do you distinguish btw. QR \& FR?]{
      \includegraphics[width=0.45\columnwidth]{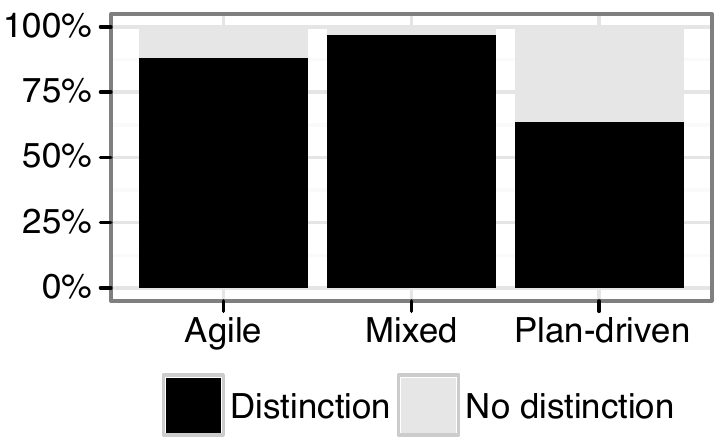}
	\label{fig:agilityvsdocumented2}
	}
   \caption{Relation between process paradigm and the style of documenting QRs.}
\label{fig:agilityvsdocumented}
\end{figure}

\noindent{\bfseries RQ1.1: Do practitioners differentiate between QRs and FRs in the
documentation? } 88\% of the respondents answered that they  document QRs in their projects, while 12\% answered that they do not document QRs at all. We contextualized this distribution w.r.t. the process paradigm the respondents use in their projects. \figurename~\ref{fig:agilityvsdocumented1} shows that all respondents with a plan-driven process document QRs, while in agile processes only 77\% document QRs.  

\begin{figure}
\centering
\includegraphics[width=0.95\textwidth]{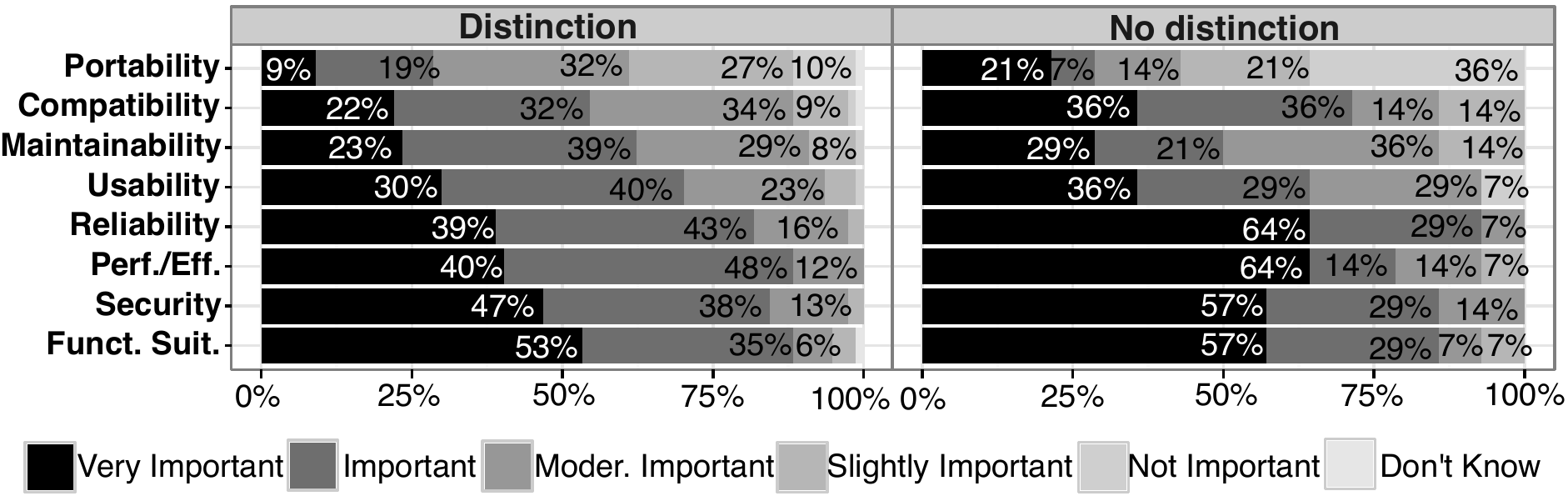}
\caption{Relation btw. importance of quality attributes and style of documentation.}
\label{fig:importancevsdistinction}
\end{figure}

From the respondents who document QRs (91 in total), 85\% answered that they distinguish between QRs and FRs in the documentation and 15\% answered that they do not. We also contextualized this distribution w.r.t. the process paradigm. \figurename~\ref{fig:agilityvsdocumented2} shows that a higher percentage of the respondents in agile processes distinguish between QRs and FRs compared with respondents in plan-driven processes. As a second contextualization, we analyzed the importance of quality factors w.r.t. the style of documentation. 
\figurename~\ref{fig:importancevsdistinction} shows how the respondents ranked the importance of different quality factors for their daily work on a five point Likert scale.
{\itshape Reliability} and {\itshape Performance\slash Efficiency}, for example, stand out as they are considered more important by participants who do not distinguish between QRs and FRs.

\begin{figure}
\centering
\includegraphics[width=0.9\textwidth]{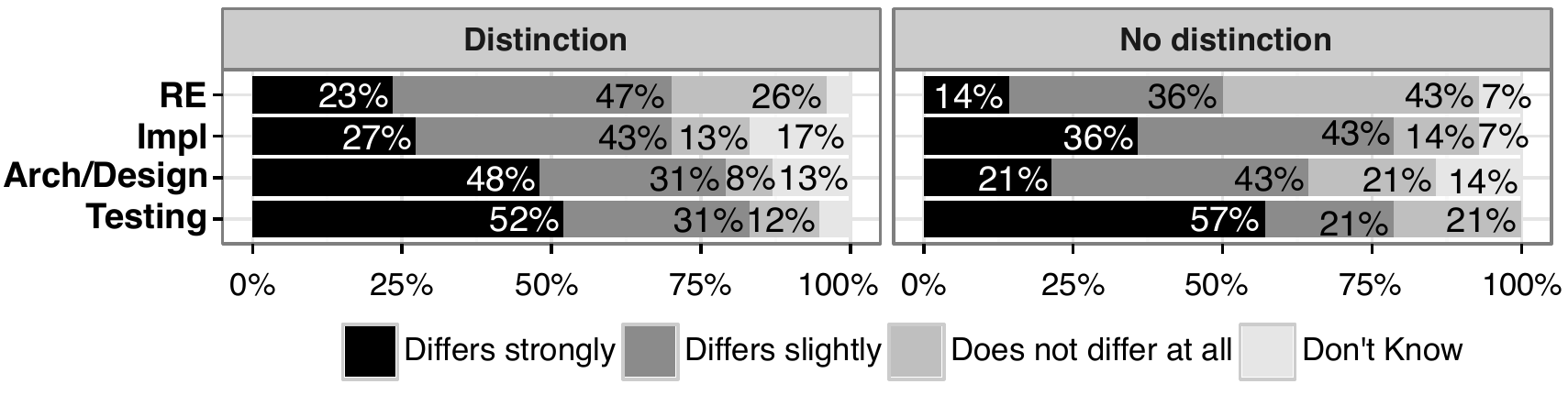}
\caption{Relation between process differences and style of documentation.} 
\label{fig:processdifferencevsdistinction}
\end{figure}

\noindent {\bfseries RQ1.2: To what extent do development activities for QRs differ from
activities for FRs? } \figurename~\ref{fig:processdifferencevsdistinction} shows how the respondents ranked the difference in the phases requirements engineering, architecture\slash design, implementation, and testing on a three point Likert scale. 
As a contextualization, we analyzed whether there is a difference in how respondents rank the difference in the development phases w.r.t. whether they do or do not distinguish between QRs and FRs.  The figure shows that the phase architecture\slash design was reported to differ stronger by respondents who distinguish between QRs and FRs.

To further detail this response, Table~\ref{tbl:differencetbl} shows exemplary statements that respondents gave explaining the differences in the development activities. 
\begin{table}
\scriptsize
\caption{Exemplary answers about differences in the development process.}
\begin{tabularx}{\columnwidth}{c l X} \toprule
{\bfseries \#}& {\bfseries Phase} & {\bfseries Answer}\\
\hline
A. &  General &{\itshape ``[QRs] are usually treated less transparent: not clearly documented, not explicitly tested, but somehow considered in RE, design and coding as common sense background, e.g., in terms of [QRs] considering IT security, performance or reliability.''} \\
B. & General &{\itshape ``FRs are documented and planned in high detail [...] Working on [QRs] are often unplanned activities and only high level documented.''}\\
C. & Test &{\itshape ``Test cases for FR[s] can quite easily [be] derived from functional models or textual requirements [... but there is no] method for deriving test cases from [QRs].''}\\
D. & Test &{\itshape ``Test planning, preparation and execution for [QRs] are handled by different stakeholders ([QRs] are [\ldots] strongly architecture related) and personnel (performance and load tests are performed by specialists usually not part of the project team).''}\\
E. & Arch. &{\itshape ``[QRs] are often architectural drivers and therefore have to be evaluated and considered very early in the project when defining the architecture. Whereas in an early stage of the project a more abstract view on the functional requirements is sufficient.''}\\
F. & Impl. &{\itshape ``[QRs] require continuous monitoring, as achievements (e.g., performance) may degrade during implementation.''}\\
G. & RE &{\itshape ``[In contrast to FRs,] [QRs] can be negotiated, if they are technically not reachable.''}\\
\bottomrule
\end{tabularx}
\label{tbl:differencetbl}
\end{table}
According to the answers, there is a different maturity of the processes for treating FRs vs. QRs (see Statement~A). Furthermore, when it comes to project planning, FRs are planned in detail but QRs are considered in an unplanned way and only documented on a high-level (see Statement~B). In testing, there are approaches for deriving test cases from FRs but none for deriving them from QRs (see Statement~C). Moreover, different stakeholders are involved in testing QRs vs. FRs (see Statement~D). In architecture and design, QRs need to be considered early in the project as they have a high impact on the architecture. In contrast to this, it is sufficient to consider FRs at an abstract level in early stages (see Statement~E). In the implementation, QRs need to be monitored continuously, whereas FRs can be implemented successively (see Statement~F). In requirements engineering, FRs are more fixed than QRs as QRs can be negotiated with the customer while FRs usually cannot (see Statement~G).



\subsection{RQ2: Reasons for Distinguishing QRs and FRs}
\begin{sidewaysfigure}
\centering
\includegraphics[width=\textwidth]{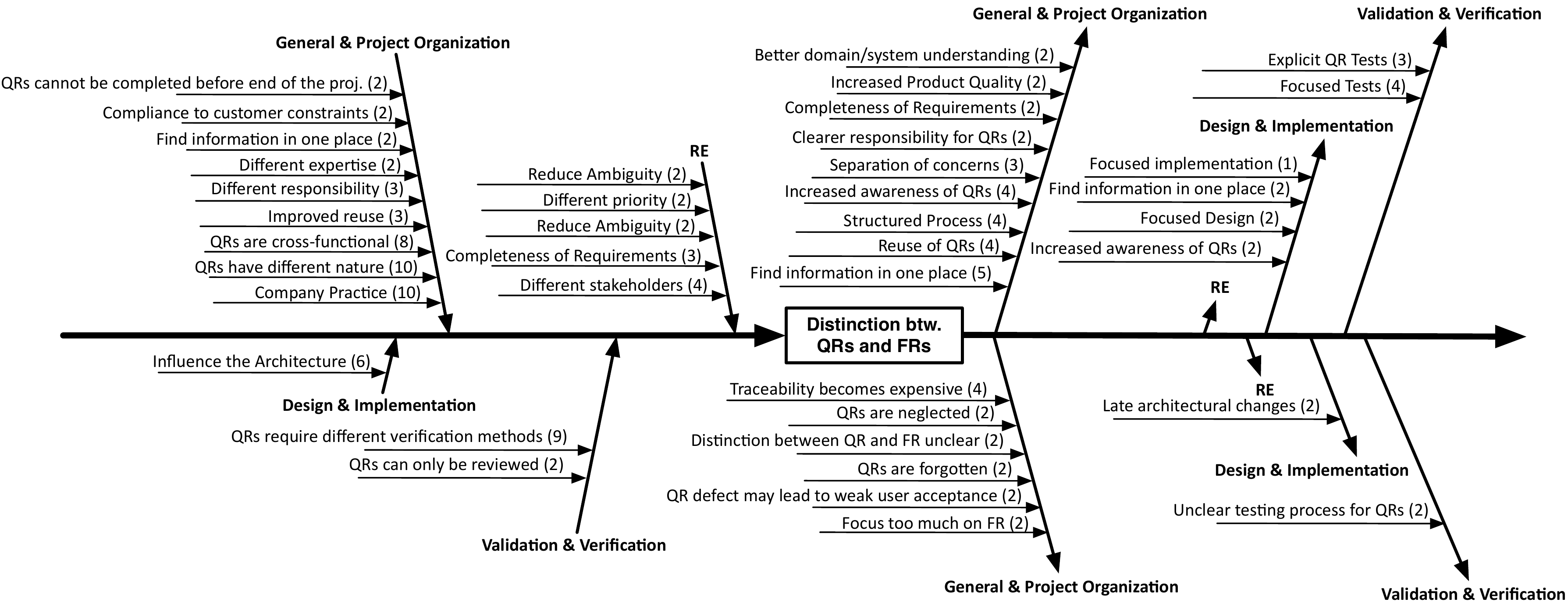}
\caption{Reasons for and consequences of distinguishing between QRs and FRs (Condensed version containing codes that occurred at least twice. The comprehensive diagram containing all codes is available at~\url{http://www4.in.tum.de/~eckharjo/DistinctionFishbone.pdf}). The left-hand side shows the mentioned reasons and the right-hand side the mentioned consequences. The upper part of the right-hand side contains the positive consequences while the lower part contains the negative consequences.}
\label{fig:distinctionfishbone}
\end{sidewaysfigure}

\begin{sidewaysfigure}
\centering
\includegraphics[width=\textwidth]{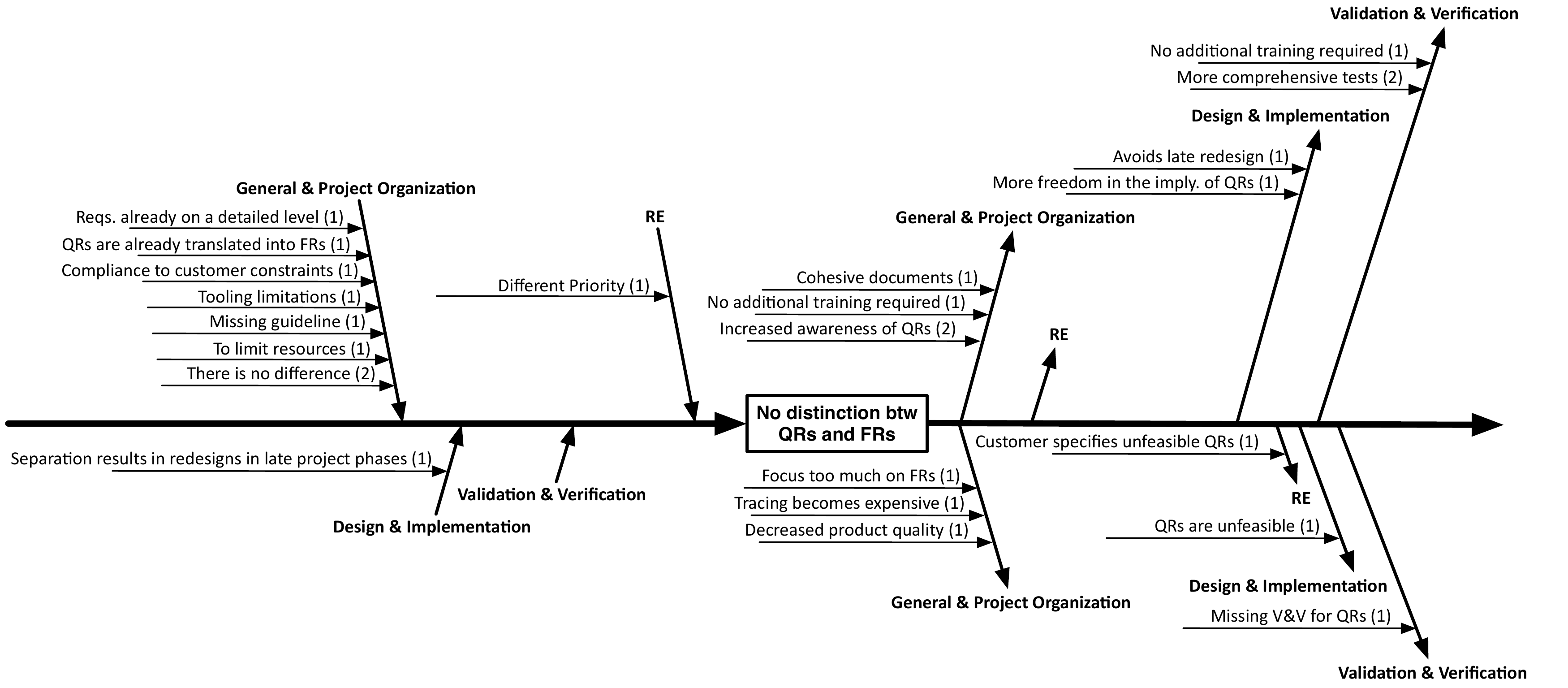}
\caption{Reasons for and consequences of not distinguishing between QRs and FRs. The left-hand side shows the mentioned reasons and the right-hand side the mentioned consequences. The upper part of the right-hand side contains the positive consequences while the lower part contains the negative consequences.}
\label{fig:nodistinctionfishbone}
\end{sidewaysfigure}

\figurename~\ref{fig:distinctionfishbone} and~\ref{fig:nodistinctionfishbone} show the cause-effect diagrams for the reasons for and consequences of (not) distinguishing between QRs and FRs in practice. On the left-hand side of the diagrams, the mentioned reasons for distinguishing (\figurename~\ref{fig:distinctionfishbone}) or not distinguishing (\figurename~\ref{fig:nodistinctionfishbone}) between QRs and FRs are indicated. 
On the right-hand side of the diagrams, the mentioned consequences of the decision are shown. The upper part contains the positive consequences while the lower part contains the negative consequences.
The different entries of the diagrams (e.g., {\itshape QRs have different nature} in \figurename~\ref{fig:distinctionfishbone}) correspond to codes that we identified in the data and their number of occurrences. Furthermore, we structured the codes in categories that are represented by the arcs in the diagram. 

\noindent {\bfseries Reasons for Distinguishing QRs and FRs: }
The left-hand side of \figurename~\ref{fig:distinctionfishbone} shows the resulting reasons for distinguishing between QRs and FRs. In total, 49 out of the 77 respondents (64\%) that distinguish between QRs and FRs provided an answer to this open question. We identified 24 codes in the answers for this question. For clarity, we only show codes that occurred at least twice in \figurename~\ref{fig:distinctionfishbone}.\footnote{The complete diagram including all codes is available at~\url{http://www4.in.tum.de/~eckharjo/DistinctionFishbone.pdf}}
Reasons that we coded as {\itshape QRs have different nature}, {\itshape Company Practice}, and {\itshape QRs are cross-functional} occur frequently in the category {\itshape General \& Project Organization}. Furthermore, in the category {\itshape Design \& Implementation} the reason {\itshape Influence the architecture} and in the category {\itshape Validation \& Verification} the reason {\itshape QRs require different verification methods} also occur often. 
 
\noindent {\bfseries Reasons for Not Distinguishing QRs and FRs: }
The left-hand side of \figurename~\ref{fig:nodistinctionfishbone} shows the mentioned reasons for not distinguishing between QRs and FRs. In total, 7 out of the 14 respondents (50\%) who do not distinguish between QRs and FRs provided an answer to this open question. We identified 8 codes in the answers for this question. \figurename~\ref{fig:distinctionfishbone} shows all identified codes, which all occurred only once in the data (except for {\itshape There is no difference}). 

%

\subsection{RQ3: Benefits and Problems}
{\bfseries Benefits and Problems of Distinguishing QRs and FRs:}
The right-hand side of \figurename~\ref{fig:distinctionfishbone} shows the consequences of distinguishing between QRs and FRs. The upper part shows the positive consequences while the lower part shows negative consequences. In total, 45 out of the 77 respondents (58\%) that distinguish between QRs and FRs provided answers to the open question about positive consequences. Regarding negative consequences, 16 out of the 77 respondents (21\%) provided answers. We identified 35 codes in the answers for positive consequences and 13 in the answers for negative consequences. 
%
As shown in the diagram, the code that we identified most in the mentioned benefits is {\itshape Find information in one place} in the category {\itshape General \& Project Organization}. In this category, there are also other benefits that occurred frequently (e.g., structuredness of the process, completeness of the requirements, separation of concerns, and increasing the awareness of QRs). We coded the benefit {\itshape Increased awareness of QRs} also three times in the category implementation. For validation and verification, the most frequent benefits are {\itshape Focused Tests} and {\itshape Explicit QRs Tests}.
%
The code that we identified most in the mentioned problems is {\itshape Traceability becomes expensive}. Further problems that were mentioned are that QRs are neglected or forgotten, that the distinction between QRs and FRs is unclear and that the distinction results in a weak user acceptance. Moreover, in the category {\itshape Validation \& Verification}, the problem {\itshape Missing testability} was mentioned.

\noindent {\bfseries Benefits and Problems of Not Distinguishing QRs and FRs:}
The right-hand side of \figurename~\ref{fig:nodistinctionfishbone} shows the consequences of not distinguishing between QRs and FRs. The upper part shows the positive consequences while the lower part shows negative consequences. In total, 9 out of the 14 respondents (64\%) that distinguish between QRs and FRs provided answers to the open question about positive consequences. Regarding negative consequences, 5 out of the 14 respondents (36\%) provided answers. We identified 7 codes in the answers for positive consequences and 6 in the answers for negative consequences.


%

\section{Discussion}
\label{sect:implications}
From the results presented in the previous section, we conclude that practitioners are split into two groups; one advocating a distinction between QRs and FRs and one advising against it. Interestingly, the respondents stated contrary reasons as arguments for or against a distinction (e.g., \emph{``Both are requirements''} vs. \emph{``We distinguish them because they are different''}). Similarly, we found the same benefits stated by respondents of both parties: 
\emph{``If you distinguish, then QRs are considered better''}
vs.
\emph{``As soon as QRs are treated equally to FRs it is a clear win-win situation such that QRs get the same attention.''}
Additionally, our results indicate that it is not clear to practitioners what the difference between both classes of requirements actually is, even though they stated reasons, benefits, and problems of a distinction: {\itshape ``Most people have problems to distinguish between them, so they mix''} or {\itshape ``[Not distinguishing] avoids unnecessary confusion at the requirements authors' side. Adding the distinction QR\slash FR would require additional training, QS, etc. without adding value to the projects''}. Some respondents see this as a reasons why they do not distinguish between them: \emph{``[\ldots]There is just no real guideline how to do it''}.

The most prevalent reasons for distinguishing between QRs and FRs are in line with those that are often found in literature (e.g., QRs have a different nature and are cross-functional, influence on architecture, require different verification methods). However, we cannot underpin any of those reasons with negative consequences in the cases where QRs and FRs were not distinguished. 
Therefore, we conclude that there seems to be confusion about this topic in practice and handling QRs seems to be driven by expectations rather than by evidence. 

In the following, we will detail and discuss some conflicting or even contradictory statements. We believe that these are topics that need to be investigated further in the future, or, in case of a clear scientific position about a topic, we need to invest more into the dissemination of the results into practice.

\noindent {\bfseries QR Testing -- A Double-edged Sword: }
One of the top reasons mentioned for distinguishing QRs and FRs was the need for different verification methods (especially w.r.t. testing). Fig.~\ref{fig:processdifferencevsdistinction} also shows that testing is the activity that differs most for QRs and FRs. When considering consequences of distinguishing between QRs and FRs in testing, we found both positive and negative. While some respondents said that a distinction leads to more focused and specialized tests for specific QRs, some also stated that a distinction leads to the fact that some QRs are not tested at all. For example, 
\emph{``Performance tests are recognized as [a] key success factor by project managers''}
vs.
\emph{``Main issue is how to handle the [QR] tests before product release''}.
On the other hand, respondents who do not distinguish between QRs and FRs also reported positive and negative consequences regarding testing:
\emph{``[\ldots]the mapping [of FRs to QRs] should ensure that this testing also covers [QRs]''}
vs.
\emph{``[When not distinguishing,] corresponding V\&V suffers''}.
We conclude from this that distinguishing QRs and FRs supports the awareness for specialized tests of important QRs but, simultaneously, bears the risk of neglecting tests for less important QRs.

\noindent {\bfseries Company Practice -- Never Change a Running Game: } 
Another commonly stated reason for distinguishing between QRs and FRs is that this is common practice in the company or that this is required by customers. However, these reasons were almost never questioned or justified. For example, 
\emph{``[\ldots]Our specification template prescribes a structuring w.r.t. [QRs] and FRs''} or
\emph{``[we distinguish] as requested by the customer''}.
Additionally, the respondents did not mention any positive or negative consequences that result from complying with customer constraints. We consider this as a sign of inadvertent handling of this topic. It would be interesting to ask customers to explicitly state reasons why they request a distinction of QRs and FRs.

\noindent {\bfseries QRs -- Drivers for the Architecture:}
Several respondents stated that the architecture of a system is specifically influenced by QRs. For example, {\itshape ``[QRs] are often architectural drivers and therefore have to be evaluated and considered very early in the project when defining the architecture''}. This was often used as an argument to distinguish between QRs and FRs: \emph{``The separation allows architects to get a quick (and in-depth) understanding of the QRs without needing to know all the functional requirements''}.
FRs, on the contrary, were considered to be more local and do not need to be fixed at the beginning of the project: \emph{``[It is] easier to find[\ldots]special FRs for developing a single use case''} or \emph{``[\ldots]in an early stage of the project a more abstract view on the functional requirements is sufficient''}.
Surprisingly, some respondents stated that it has a positive impact for the implementation when QRs and FRs are not strictly distinguished: {\itshape ``[QRs] and FRs are handled as features. They are not separated, which avoids the redesigns e.g., due to performance problems''} and {\itshape ``[When not distinguishing,] we have much more freedom during the implementation iterations[\ldots]to find solutions that fit the customers' expectations and the possibilities that come with the architecture and technology we use''}.

\noindent {\bfseries Awareness Matters: }
It seems that an increased awareness for QRs was considered as one of the most prominent benefits. Both parties claimed this as a benefit of distinguishing respectively not distinguishing between QRs and FRs: {\itshape ``[Distinction] ensures that [QRs] are also in the focus''} vs. {\itshape ``[Not distinguishing] helps keeping the team aware that the device does not only need to have certain features, but that these features also need to work e.g., at a high temperature''}.
It seems that awareness can be increased with both strategies. The crucial point seems to be that there is a clear and explicit relation between FRs and QRs, which leads to the following observation.

\noindent {\bfseries Tracing -- The Good, the Bad, and the Ugly: }
One trade-off that we found in the data is an inherent challenge that does not seem to be resolved in practice. Some respondents stated that a distinction between QRs and FRs is beneficial because it keeps associated information in one place and, thus, supports different viewpoints on the requirements: {\itshape ``People who are particularly concerned with QRs, such as architects and performance testers, find relevant information in one place''} and {\itshape ``As most [QRs] apply across components, they are more easily retrieved in a separate specification''}.
However, this benefit also comes with clear disadvantages considering tracing and the risk of forgetting requirements: {\itshape ``Consistent documentation of relationships between FRs and [QRs] is difficult''} and {\itshape ``The development team needs to be fully aware about all sources for requirements. Ostrich strategy causes a high yield of trouble''}.
Respondents who do not distinguish reported on benefits regarding the cohesiveness of their specifications: {\itshape ``Some documents benefit from this, as they turn more cohesive''} or {\itshape ``[\ldots]the feature is really ready if installed and not only 80\%''}.

\section{Limitations and Threats to Validity}
\label{sect:threats}
We now discuss the threats to validity and mitigation measures we applied. 

\noindent {\bfseries Participant Selection: }
One limitation in the study is the missing lack of control over the respondents given that we distributed the survey invitation over various networks. Apart from an unknown response rate, this means that we cannot control how representative the responses are. We removed those respondents from the population that stated that they do not deal with requirements. Also, although the introductory texts explicitly stated that the survey is aimed at addressing practitioners perspective, we cannot guarantee that all the views taken really result from practitioners. 

\noindent {\bfseries Survey Research:}
Further threats to the validity result from the nature of survey research. We cannot control on which basis the respondents provide their answers, the respondents might be biased, and there exists the possibility that they have misinterpreted some of the questions or even the concept of QR/NFR. We reduced the first threat by asking questions to characterize the context of the respondents. We cannot mitigate the second threat, but reduced it by conducting the survey anonymously. We minimized the third threat by conducting a pilot phase in which we tested the instrument used and the data analysis techniques applied. 

\noindent {\bfseries Subjectivity of Coding:}
A further major threat to validity, however, arises from the data analysis, i.e., the coding process, because coding is a creative task. Subjective views of the coders, such as experiences and expectations, might have influenced the way we coded the free text statements. A threat arises from the fact that we cannot validate our results with the respondents given the anonymous nature of our survey. We minimized this threat by coding in pairs (researcher triangulation). 

\noindent {\bfseries Representativeness of the Codes:}
Finally, one limitation stems from the result set itself and its expressiveness. Our focus was to collect and code practitioners experiences on how they consider QRs. We quantified the results to get an overview of whether certain codes dominate others. However, a potentially high frequency of codes does still not allow for conclusions on the criticality of those codes. In particular, the fact that we got more answers about reason for and consequences of a distinction between QRs and FR than for no distinction might have distorted our interpretation of the results. 

\section{Related Work}
\label{sect:rw}

The literature on categorizations of requirements is very extensive. Major contributions address categorizing non-functional requirements (e.g.,~\cite{chung2009non,Glinz07,Pohl10}), of which most rely on quality (definition) models (a detailed discussion can be found in~\cite{Eckhardt16}). Pohl~\cite{Pohl10}, for instance, discusses the misleading use of the term ``non-functional'' and argues to use ``quality requirements'' for product-related NFRs that are not constraints.
Glinz~\cite{Glinz07} performs a comprehensive review on the existing definitions of NFRs, analyzes problems with these definitions, and proposes a definition on his own. Mairiza~et~al.~\cite{mairiza2010investigation} perform a literature review on QRs, investigating the notion of QRs in the software engineering literature to increase the understanding of this complex and multifaceted phenomenon. They found 114 different QR classes. Contributions such as those have fostered valuable discussions on the fuzzy terminology used and the concepts applied, but they did not focus on the implications of these categorizations on development processes in practice. \looseness=-1


Chung and Nixon~\cite{chung1995dealing} investigate how practitioners handle QRs. They argue that QRs are often retrofitted in the development process or pursued in parallel with, but separately from, functional design and that an ad hoc development process often makes it hard to detect defects early. They perform three experimental studies on how well a given framework~\cite{Mylopoulos92} can be used to systematically deal with QRs. Svensson~et~al.~\cite{svensson2009quality} perform an interview study on how QRs are used in practice. Based on their interviews, they found that there is no QR-specific elicitation, documentation, and analysis, that QRs are often not quantified and, thus, difficult to test, and that there is only an implicit management of QRs with little or no consequence analysis. Furthermore, they found that at the project level, QRs are not taken into consideration during product planning (and are thereby not included as hard requirements in the projects) and they conclude that the realization of QRs is a reactive rather than proactive effort. 

Borg et al.~\cite{borg2003bad} analyze via interviews how QRs are handled in two Swedish software development organizations. They found that QRs are difficult to elicit because of a focus on FRs, they are often described vaguely, are often not sufficiently considered and prioritized, and they are sometimes even ignored. Furthermore, they state that most types of QRs are difficult to test properly due to their nature, and when expressed in non-measurable terms, testing is time-consuming or even impossible. Ameller~et~al.~\cite{Ameller12} perform an empirical study based on interviews around the question {\itshape How do software architects deal with QRs in practice?} They found that QRs were often not documented, and even when documented, the documentation was not always precise and usually became desynchronized. 

In all of the investigations, FRs and QRs are treated separately, and the investigations take an observational perspective on how practitioners deal with QRs in that context. The goal of our study is to analyze whether practitioners handle FRs and QRs differently, which reasons motivate the way they consider QRs, and what consequences this has on the development process.

\section{Conclusions}
\label{sect:conclusion}
In this paper, we reported on a survey conducted with 103 practitioners to explore whether and, if so, why they handle requirements labeled as ``functional'' differently from those labeled as ``quality'' as well as to disclose resulting consequences for the development process. Our results indicate that practitioners document QRs and most of them do make an explicit distinction between QRs and FRs in the documentation. Furthermore, our data suggests that the development process strongly differs depending on a distinction between QRs and FRs, especially in interconnected activities such as testing. The rationale of practitioners is that QRs are different to FRs, i.e. they are of different nature, are cross-functional, strongly influence the architecture, and require different verification methods. In our previous study~\cite{Eckhardt16}, we found, however, that many requirements labeled as ``quality'' might as well be categorized as ``functional'' and prior to the study presented here, we had the simple speculation that if a blurry distinction determines how the following development activities are performed, we should find problems that arise because the activities do not really fit the corresponding requirements. Still, our results indicate that the question whether to make a distinction or not is without a direct linkage to negative or positive consequences per se. Therefore, we argue that the decision whether to make an explicit distinction should be made consciously such that people are aware of the risks that this distinction bears so that they may take countermeasures. 

{\bfseries Acknowledgements:}
We would like to thank M. Broy, K. Beckers, J. Mund, S. Smith-Eckhardt, and M. Glinz for their helpful comments and suggestions.

\bibliographystyle{splncs03}
\bibliography{bib}
\end{document}